\documentclass[%
reprint,
amsmath,amssymb,
aps,
]{revtex4-2}
\usepackage{xcolor}
\usepackage{graphicx}
\usepackage{dcolumn}
\usepackage{lipsum}
\usepackage{bm}
\usepackage{xcolor}
\usepackage{hyperref}
\usepackage{tikz}
\usepackage{orcidlink}
\usepackage{subfig}

\usepackage{physics}
\usepackage{appendix}
\usepackage{orcidlink}
\begin{document}
	\preprint{APS/123-QED}
	\preprint{APS/123-QED}
	\title{Statistical analysis of quantum trajectories in dissipative Landau-Zener model}
	\author{Laleh Memarzadeh~\orcidlink{0000-0002-3394-4214}
	}
	\affiliation{%
		Department of Physics, Sharif University of Technology, Tehran 11155-9161, Iran}
	\author{Rosario Fazio~\orcidlink{0000-0002-7793-179X}
	}
	\affiliation{The Abdus Salam International Center for Theoretical Physics, Strada Costiera 11, 34151 Trieste, Italy}
	\affiliation{Dipartimento di Fisica ``E. Pancini", Universit\`a di Napoli ``Federico II'', Monte S. Angelo, I-80126 Napoli, Italy}
	
	\date{\today}
	
	\begin{abstract}
		We present statistics of quantum jumps in the two-level system with landau-Zener Hamiltonian that undergoes a Markovian process. For the Landau-Zener model, which is successful in simulating adiabatic/non-adiabatic evolution and quantum annealing, we consider two types of dissipation. In the first one, the jump operators project states to the initial ground state and excited state of the Hamiltonian at $t\to -\infty$. In the second type, the jump operators project to the instantaneous eigenstates of the Hamiltonian. By the quantum trajectories approach, we present the probability of the number of jumps in adiabatic and non-adiabatic regimes for both models. 
		Furthermore, we demonstrate the statistics of jumps in time intervals of the evolutions. Also, we show the role of bath temperature, coupling strength to the environment, and spin-coupling directions on the statistics of quantum jumps. 
	\end{abstract}
	\maketitle
	\section{Introduction}
	Recent developments in quantum information science and technology have led to successful implementation of different quantum protocols. In all quantum protocols, the realistic approach is to consider the presence of quantum noise due to the unavoidable interaction between the system of interest and its surrounding environment. 
	Hence, the initial pure state evolves to a mixed state described by a density operator which is a probabilistic mixture of pure states in time. 
	Although a density operator provides a very informative description of the system in time, it does not reflect abrupt stochastic transitions during the evolution of each pure state. 
	
	The probabilistic transitions in each pure state have a significant practical impact on quantum computation. The most remarkable instances are adiabatic quantum computation \cite{Farhi2000, Aharonov2007} and quantum annealing \cite{Santoro2006, Crosson2021}.
	The two-level Landau-Zener model is a fine approximation for the lowest energy level of many-body interacting systems \cite{Santoro2002} and allows us to study the adiabatic/non-adiabatic evolution \cite{Dziarmaga2010, Polkovnikov2011} which is especially important in adiabatic quantum computation \cite{Farhi2000, Aharonov2007} and quantum annealing \cite{Santoro2006, Crosson2021}. The dissipative Landau-Zener model has been realized experimentally \cite{Zenesini2009, Dickson2013, Troiani2017, Zhang2018, Zhu2021} and has been studied from different aspects including quantum annealing \cite{Arceci2017}, transition probabilities \cite{Gefen1987,Shimshoni1991, Saito2007, Ashhab2014, Ashhab2016, Arceci2017, Kayanuma1998, Ao1991, Wubs2006, Zueco2008, Orth2010, Nalbach2010, Huang2018, Nalbach2022, Zhang2023, Ashhab2023,  Zhang2024}, entanglement degradation \cite{Babakan2023}, realization of universal quantum gates \cite{Wang2016} and adiabatic quantum computation \cite{Ashhab2006}. 
	
	To go beyond the average description of the system given by a density operator, we employ the quantum trajectory \cite{Dum1992, Dalibard1992, Molmer1993, PlenioKnight1998, Daley2014} approach to obtain statistical information about the system evolution. The quantum trajectory approach is based on monitoring the system in time through measurements on the environment \cite{PlenioKnight1998}. In this construction, one obtains a stochastic description of the evolution of any individual pure state. Therefore, this formalism not only enriches our understanding of the evolution of quantum systems but also provides the basis for evaluating the success of a single realization of any given quantum protocol. Furthermore, this construction supplies an algorithmic approach for solving master equations. The agreement between the exact results and results presented by the quantum trajectories, within the statistical errors, shows the success of this method for describing the dynamics of open quantum systems \cite{Daley2014}. 
	Quantum trajectories have been used for different goals such as phase transition detection \cite{GarrahanLesanovsky2010}, quantum synchronization \cite{Najmeh2020}, quantum control \cite{Lange2014}, 
	entanglement generation \cite{Chantasri2016} and environment assisted quantum state transfer \cite{King2024}. 
	In parallel experimental progress has made it possible to observe quantum trajectories \cite{Roch2014, Vool2014} and construct the foundation for further applications of quantum trajectories. 
	
	Here, we aim to employ the quantum trajectory approach to deliver statistical information about the system evolution that can not be captured by studying its density operators. 
	For that, we consider
	a two level system with the Landau-Zener Hamiltonian \cite{Landau1932,Zener1932} that interacts with a thermal bath.
	We focus on two different types of dissipative Landau-Zener models. For each model, we use the quantum Monte Carlo algorithm to simulate the dynamics and study the quantum trajectories. 
	We present the probability of having quantum jumps due to the interaction with the environment both in adiabatic and non-adiabatic regimes. Furthermore, we discuss the statistics of having quantum jumps in different time intervals during the evolution. 

	We begin in \S\ref{sec:BackGround} with a review of the Monte Carlo method for simulating the dynamics of open quantum systems, which is based on the idea of quantum trajectories. 
 In \S\ref{sec:Model} we introduce two different dissipative dynamics for a two-level system with Landau-Zener Hamiltonian. Our results about the statistics of quantum jumps are presented in \S\ref{sec:statistics}. \S\ref{sec:Conclusion} contains  the discussion of the results and conclusion.
	
	\section{Background}\label{sec:BackGround}
	In this section, we review 
	quantum trajectory approach. This approach was developed in \cite{Molmer1993} as a Monte Carlo wave-function method for simulating the dynamics of open quantum systems. 
	
	For an open quantum system, the general form of Hamiltonian is given by
	\begin{equation}
		H_{\text{total}}=H_{\rm S} + H_{\rm E}+\lambda H_{\rm SE}.
	\end{equation}
	The Hamiltonian of the system, environment and interaction between the system and the environment are denoted by $H_S$ $H_E$ and $H_{SE}$. The real parameter $\lambda$ denotes the coupling constant between the system and the environment. In the weak coupling limit, the Markovian master equation \cite{Gorini1976, Lindblad1976} for system evolution is given by	\begin{align}\label{eq:GKLS}
		\dot{\rho}&=-i[H_{\rm S},\rho]\cr
		&+\lambda^2\sum_{m}\gamma_m(A_m\rho A_m^\dagger-\frac{1}{2}\{A_m^\dagger A_m,\rho\})
	\end{align}
	in which $\rho$ is the density operator of system, $\gamma_m>0$ are Kossakowski coefficients, $A_m$'s are Lindblad operators \cite{Gorini1976, Lindblad1976}. For proceeding discussions it is convenient to define jump operators $C_m:=\sqrt{\gamma_m}A_m$. 
	
	Solution of the master equation in Eq.~(\ref{eq:GKLS}) provides the density matrix of the system at an arbitrary time. That is while, the Monte Carlo wave-function method, enables us to study every individual quantum trajectory described by a pure state. 
	In this method by having $\ket{\phi(t)}$, which is the state of the system at time $t$, one can compute the state of the system at time $t+\delta t$, denoted by $\ket{\phi(t+\delta t)}$, by considering the probability of different possible transitions which are described by Lindblad operators in Eq~(\ref{eq:GKLS}). More specifically, a non-Hermitian Hamiltonian $H$ is defined in terms of $H_S$, and the jump operators $C_m$ as follows:
	\begin{equation}
		H=H_{\rm S}-\frac{i\lambda^2}{2}\sum_{m=1}^M C_m^\dagger C_m. 
	\end{equation}
	Then comparison between a randomly generated number $\epsilon\in[0,1]$ and $\delta p$
	\begin{equation}
		\delta p=i\delta t \bra{\phi(t)}|H-H^\dagger\ket{\phi(t)},
	\end{equation}
	determines if there is a jump or not. If $\delta p<\epsilon$ then there is no quantum jump and the state of the system at time $t+\delta t$ is given by
	\begin{equation}
		\ket{\phi(t+\delta t)}=\frac{(\operatorname{id}-i H\delta t)}{\sqrt{1-\delta p}}\ket{\phi(t)}.
	\end{equation}
	If $\delta p\geq\epsilon$, then there is a quantum jump with probability $\delta p_m$
	\begin{equation}
		\delta p_m=\frac{\lambda^2\delta t}{\delta p}\bra{\phi(t)}C_m^\dagger C_m\ket{\phi(t)},
	\end{equation}
	
	and the 
	state at time $t+\delta t$ is given by
	\begin{equation}
		\ket{\phi(t+\delta t)}=\lambda\sqrt{\frac{\delta t}{\delta p\delta p_m}}C_m\ket{\phi(t)}.
	\end{equation}
	It is shown that by running the algorithm several times, all of them starting from the same initial state at time $t_{\rm int}$, the average over all possible events (different jumps or evolving without any jump), is equal to the density matrix description given by the master equation in Eq.~(\ref{eq:GKLS}) \cite{Daley2014}.
	
	\section{Model }\label{sec:Model}
	In this section, we consider a two-level Landau-Zener system in interaction with a thermal bath. After recalling the original Landau-Zener model, we focus on two types of dissipative Landau-Zener models described by a quantum Markov process. 
	
	The  time-dependent Landau-Zener Hamiltonian for a qubit system is given by
	\begin{equation}
		\label{eq:LZ-Hamiltonian}
		H_{\text{LZ}}=vt\sigma_z+\Delta\sigma_x,
	\end{equation}
	where $\sigma_x$ and $\sigma_z$ are the well-known Pauli operators, $\Delta$ and $v$ are positive real parameters, with $v$ describing the driving velocity and $\Delta$ denoting the strength of magnetic field along $x$ direction.  We denote the instantaneous eigenstate of $H_{\text{LZ}}$ with $\ket{\epsilon_{\pm}(t)}$:
	\begin{equation}
		\label{eq:LZHamiltonian}
		H_{\rm{LZ}}\ket{\epsilon_\pm(t)}=\epsilon_\pm(t)\ket{\epsilon_{\pm}(t)}
	\end{equation}
	where
	\begin{align}
		&\epsilon_+(t)=-\epsilon_-(t)=\sqrt{v^2t^2+\Delta^2}\cr\cr
		&  \ket{\epsilon_\pm}=\frac{\Delta\ket{e}+(\epsilon_\pm(t)-vt)\ket{g}}{\sqrt{2\epsilon_\pm(t)(\epsilon_\pm(t)-vt)}},
	\end{align}
	where $\ket{e}$ and $\ket{g}$ are eigenstates of $\sigma_z$ with eigenvalues $1$ and $-1$.
	In the adiabatic regime which is characterized by $\frac{\Delta^2}{v}\gg 1$, there is 
	no transition between the eigenstates of the system. That is if at $t\to -\infty$ system is in its ground state or excited state, namely $\ket{\epsilon_{\pm}(-\infty)}$ it remains there during the evolution and described by $\ket{\epsilon_{\pm}(t)}$ till the end of evolution at $t\to\infty$. Contrary to the adiabatic regime, in the non-adiabatic regime characterized by $\frac{\Delta^2}{v}\ll 1$, there are transitions between eigenstates of the Hamiltonian. 
	Adiabatic and non-adiabatic evolution can be distinguished by studying the probability of finding the system at its ground state at $t\to\infty$ which is $\ket{g}$ when the system is initially in its ground state at $t\to -\infty$ which is $\ket{e}$. Landau and Zener reported an analytical expression for this quantity, called $P_{\ket{e}\to\ket{g}}$, in terms of $\Delta$ and $v$ \cite{Landau1932, Zener1932}. Large values of $P_{\ket{e}\to\ket{g}}$ indicate adiabatic evolution and small values of $P_{\ket{e}\to\ket{g}}$ demonstrate the occurrence of transitions between instantaneous eigenstates of the Hamiltonian $H_{\rm LZ}$ and hence is a sign of a non-adiabatic evolution. 
 Analyzing transition probabilities in the Landau-Zener has been extended to non-reversible or dissipative dynamics. In \cite{ Gefen1987,Shimshoni1991}, the interaction between the qubit system and a phonon bath has been considered to discuss the role of the bath's temperature, relaxation and dephasing on the transitions and the system's behaviour. These studies have been extended to more general settings and discussions on the impact of different prameters on the transition probabilities \cite{Saito2007, Arceci2017, Kayanuma1998, Ao1991, Wubs2006, Zueco2008, Orth2010, Nalbach2010, Huang2018, Nalbach2022, Zhang2023,  Zhang2024}. 

	In the following, we consider two different dissipative Landau-Zener models.
	\subsection{Type I}
	In this subsection, we present the first type of dissipative Landau-Zener model. We introduce the master equation and discuss its action on the states. 
	
	In this model, the master equation for system density operator $\rho(t)$ is given by
	\begin{align}
		\dot{\rho}=\label{eq:SimpleM}
		\mathcal{L}_{\rm I}[\rho]=-i[H_{\rm{\rm LZ}},\rho]
		&+\gamma(1+\tau)(\sigma_-\rho\sigma_+-\frac{1}{2}\{\sigma_+\sigma_-,\rho\})\cr
		&+\gamma\tau (\sigma_+\rho\sigma_- -\frac{1}{2}\{\sigma_-\sigma_+,\rho\}).
	\end{align}
	Here, $\tau$ is a non-negative real parameter miming the environment's temperature. For zero-temperature environment $\tau=0$. The strength of interaction with the environment is characterized by positive parameter $\gamma$. 
	In this model, we have two Lindblad operators
	\begin{align}
		\label{eq:JumpTI}
		& C^{(\rm I)}_1:=\sqrt{(1+\tau)}\sigma_-=\sqrt{(1+\tau)}\ket{g}\bra{e}\cr
		& C^{(\rm I)}_2:=\sqrt{\tau}\sigma_+=\sqrt{\tau}\ket{e}\bra{g}.
	\end{align}
	These jump operators project each state to the eigenstates of the system's Hamiltonian at time $t\to -\infty$ that are the eigenstates of the spin operator in $z$ direction. 
	\subsection{Type II}
	In this model, we consider a particular form of interaction between the system and environment given by
	\begin{equation}
		\label{eq:Hint}
		H_{\rm{int}}=\boldsymbol{A}\otimes \int_0^{\omega_{\rm max}}d\nu h(\nu)(b(\nu)+b^{\dagger}(\nu)).
	\end{equation}
	with
	\begin{equation}
		\boldsymbol{A}=\frac{1}{2}(\cos(\theta)\sigma_z+\sin(\theta)\sigma_x)
	\end{equation}
	Following the procedure of deriving the master equation in the weak coupling limit, the master equation is given by \cite{Arceci2017, Yamaguchi2017, Babakan2023} 
	\begin{align}
		\label{eq:Master2}
		\dot{\rho}&=-i\mathcal{L}_{\rm II}[\rho]=-i[H_{\rm LZ},\rho]\cr
		&+\lambda^2\sum_{k=1}^3\gamma_k (A_k\rho A_k^\dagger-\frac{1}{2}\{A_k^\dagger A_k,\rho\})
	\end{align}
	In this master equation, Lindblad operators project to the instantaneous eigenstates of the Hamiltonian $H_{\rm LZ}$:
	\begin{align}
		\label{eq:lindbladII}
		&A_1=a_1(t) \ket{\epsilon_-(t)}\bra{\epsilon_+(t)}\cr
		&A_2=a_2(t)\ket{\epsilon_+(t)}\bra{\epsilon_-(t)}\cr
		&A_3=a_3(t)\ket{\epsilon_-(t)}\bra{\epsilon_-(t)}+a'_3(t)\ket{\epsilon_+(t)}\bra{\epsilon_+(t)}\cr
	\end{align}
	The time-dependent coefficients $a_i$ are given in terms of $\boldsymbol{A}$ and depend on the spin-coupling direction $\theta$:
	\begin{align}
		&a_1(t)=a_2^*(t)=\bra{\epsilon_-(t)}\boldsymbol{A}\ket{\epsilon_+(t)}\cr
		&a_3(t)=\bra{\epsilon_-(t)}\boldsymbol{A}\ket{\epsilon_-}\cr
		&a'_3(t)=\bra{\epsilon_+(t)}\boldsymbol{A}\ket{\epsilon_+}.
	\end{align}
	In Eq.~(\ref{eq:Master2}), positive coefficients $\gamma_m$ are given by
	\begin{align}
		\label{eq:GammaII}
		&\gamma_1=2\pi J (n+1)\cr
		&\gamma_2=2\pi J n\cr
		&\gamma_3=2\pi T
	\end{align}
	where $T$ is the bath temperature, $n$ is the mean photon number in the bath
	\begin{equation}
		n=\frac{1}{e^{\frac{2\epsilon_+(t)}{T}}-1}
	\end{equation}
	and $J$ is the spectral density
	\begin{equation}
		J=2\epsilon_+(t)e^{\frac{2\epsilon_+(t)}{\omega_c}}
	\end{equation}
	$\omega_c$ is a positive parameter which takes values such that the approximations for constructing the time-dependent generator in Eq.~(\ref{eq:Master2}) \cite{Yamaguchi2017, Babakan2023} are valid. Therefore, the ultimate from of jump operators are given by
	\begin{equation}
		\label{eq:JumpGLB}
		C_m^{(\rm II)}=\sqrt{\gamma_m}A_m,\;\;m=1,2,3
	\end{equation}
	with $A_m$ given in Eq.~(\ref{eq:lindbladII}) and $\gamma_m$ in Eq.~(\ref{eq:GammaII})
	 parameters
	\begin{figure}
		\centering
		\includegraphics[width=\columnwidth]{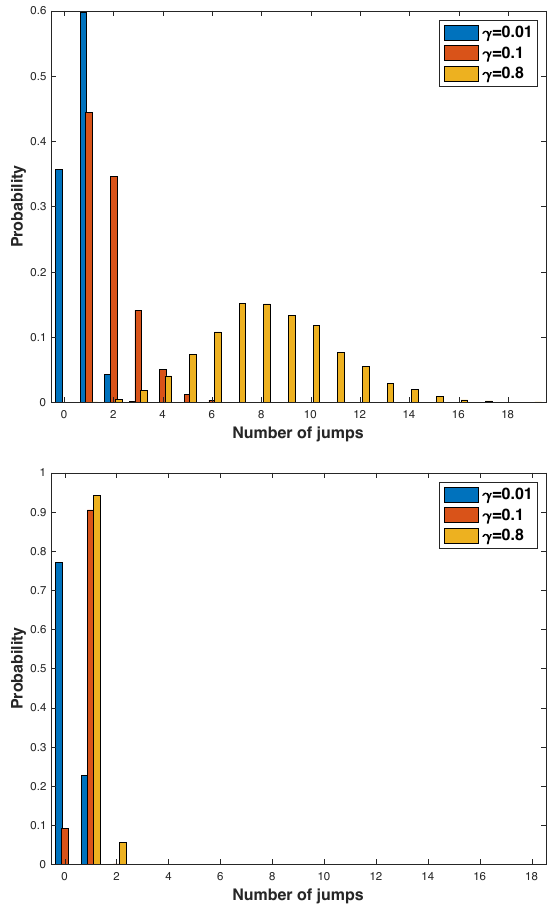}
		\caption{Type I: Probability of numbers of jumps for different values of $\gamma$, when $\Delta=1$, $\tau=0$,  $v=0.1$ (Top)  and  $v=10$ (bottom).}
		\label{fig:T1_PvsNDiffGamma}
	\end{figure}	
   \begin{figure}
   	\centering
   	\includegraphics[width=\columnwidth]{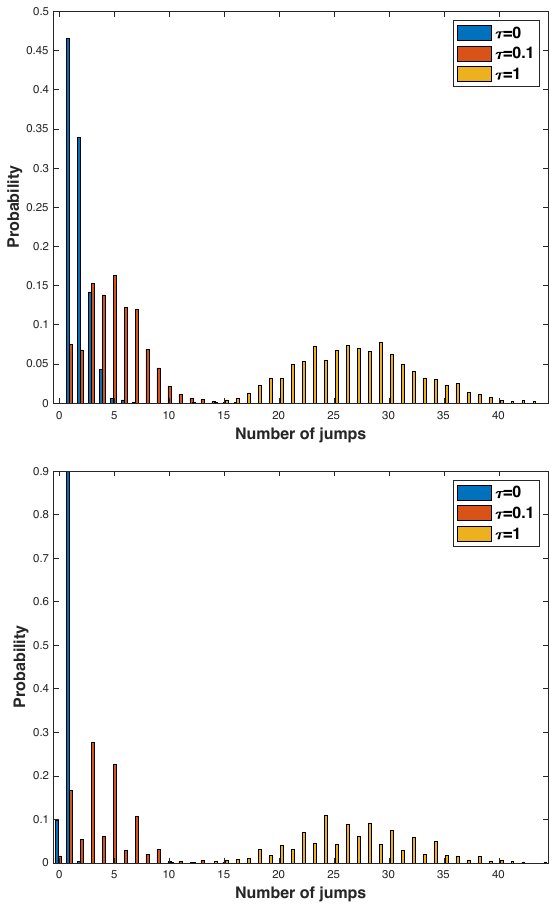}
   		\caption{Type I: Probability of numbers of jumps for different values of $\tau$, when $\Delta=1$, $\gamma=0.1$ and $v=0.1$ (Top), $v=10$  (bottom).}
   	\label{fig:T1_PvsNDiffTau}
   \end{figure}
		

	\begin{figure}
		\centering
		\includegraphics[width=\columnwidth]{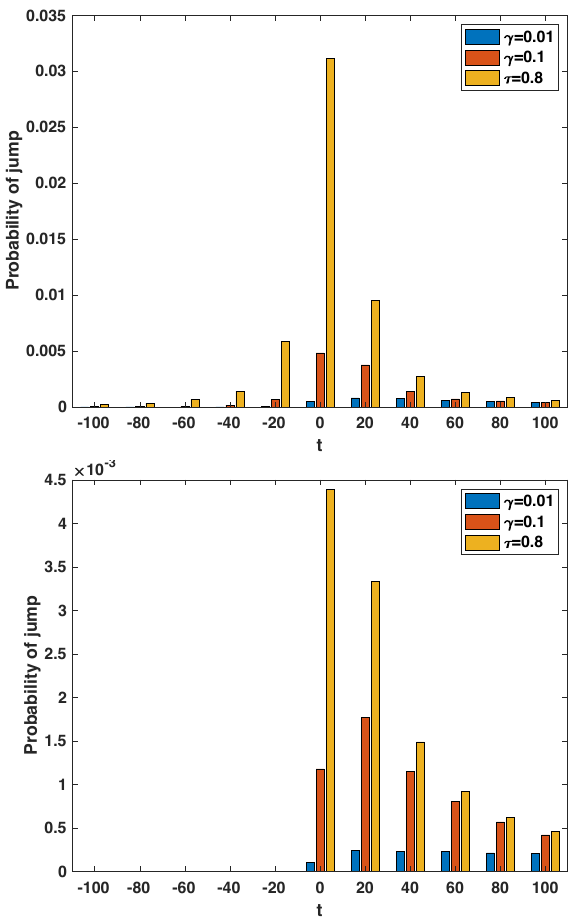}
			\caption{Type I: Probability of jumps in time intervals $\Delta t=20$ for different values of $\gamma$, when $\Delta=1$, $\tau=0$.  $v=0.1$ (Top)  $v=10$ (bottom).}
		\label{fig:LZ1IntervalsTau0}
	\end{figure}
	\begin{figure}
		\centering
		\includegraphics[width=\columnwidth]{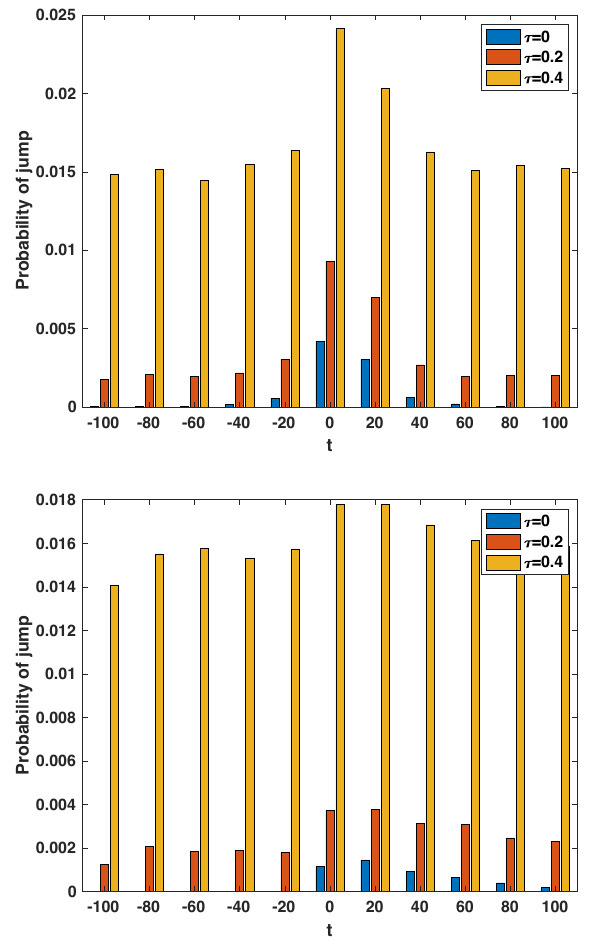}
			\caption{Type I: Probability of jumps in time intervals $\Delta t=20$ for different values of $\tau$, when $\Delta=1$, $\gamma=0.1$.  (Top) $v=0.1$ (bottom) $v=10$.}
		\label{fig:LZ1IntervalsGamma01}
	\end{figure}

	\begin{figure}
		\centering
		\includegraphics[width=0.9\columnwidth]{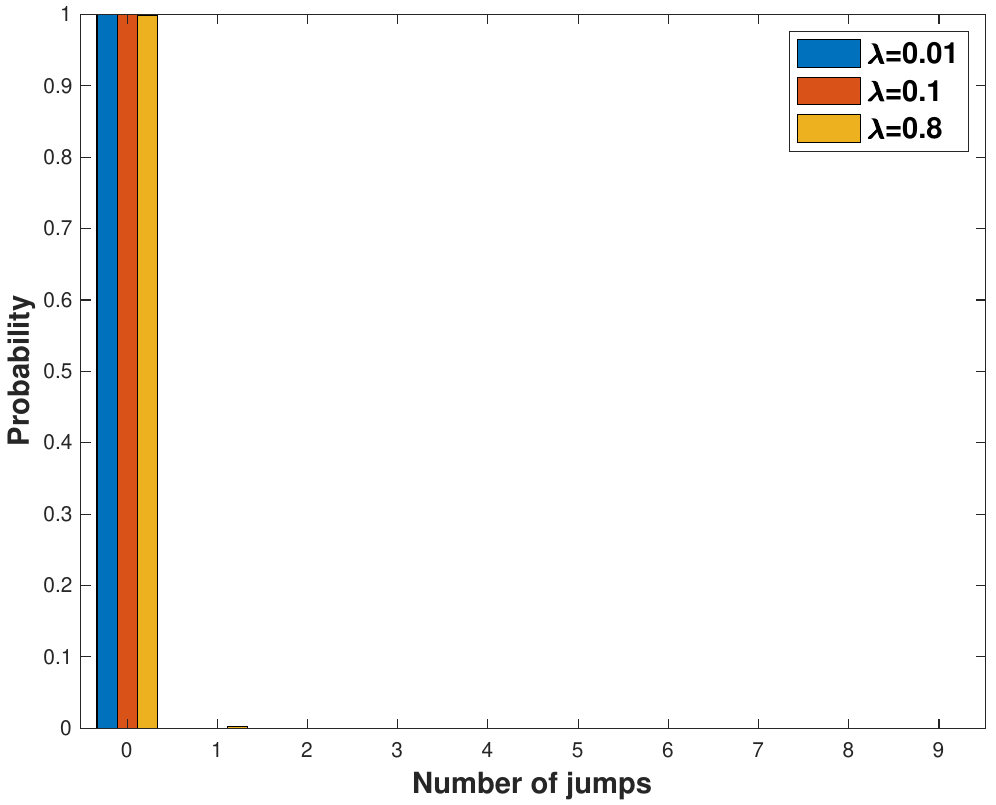}
		\caption{Type II: Probability of jumps for different values of $\lambda$, when $\Delta=1$, $T=0$, $\theta=0$ and $v=0.1$}
		\label{fig:T2_PvsNDiffTGammaT0Theta0V01}
	\end{figure}
	\begin{figure}
		\centering
		\includegraphics[width=\columnwidth]{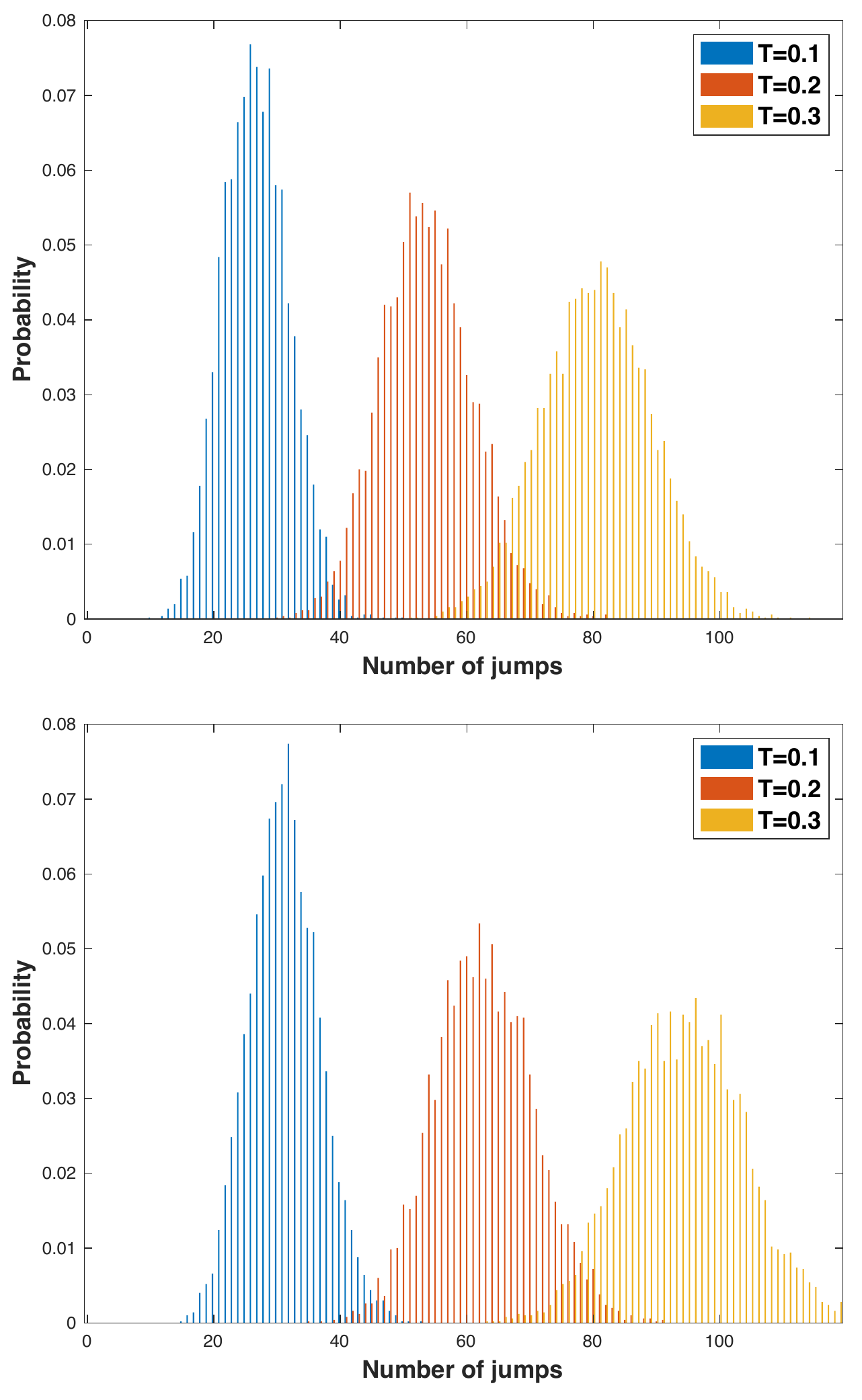}
			\caption{Type II: Probability of jumps for different values of $T$, when $\Delta=1$, $\gamma=1$, $\theta=0$ and $v=0.1$ (Top)  $v=10$ (bottom).}
		\label{fig:T2_PvsNDiffTTheta0}
	\end{figure}
	\begin{figure}
		\centering
		\includegraphics[width=\columnwidth]{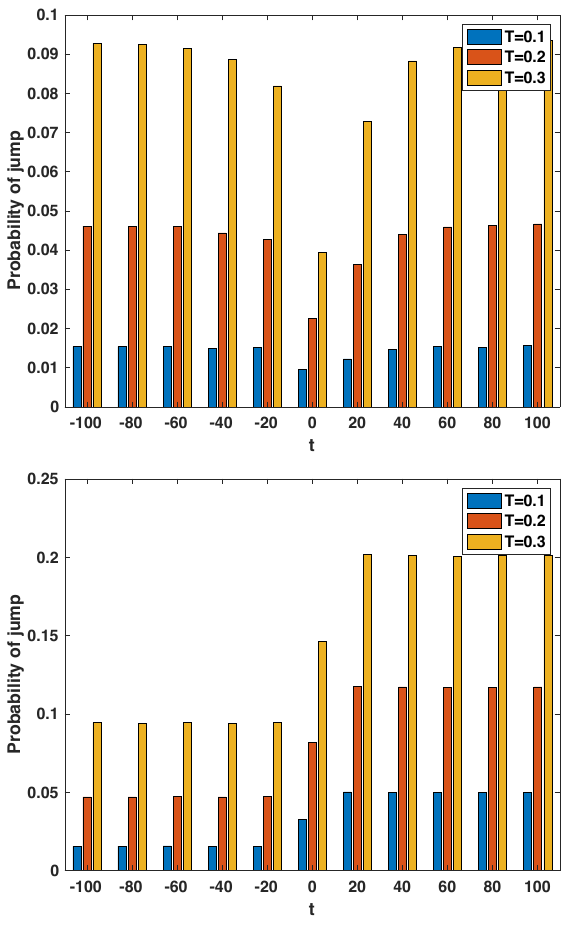}
			\caption{Type II: Probability of jumps in time intervals $\Delta t=20$ for different values of $T$, when $\Delta=1$, $\gamma=1$ and $\theta=0$ and  $v=0.1$ (Top), $v=10$ (bottom).}
		\label{fig:LZ2IntervalsGamma1Theta0}
	\end{figure}

	\begin{figure}
		\centering
		\includegraphics[width=\columnwidth]{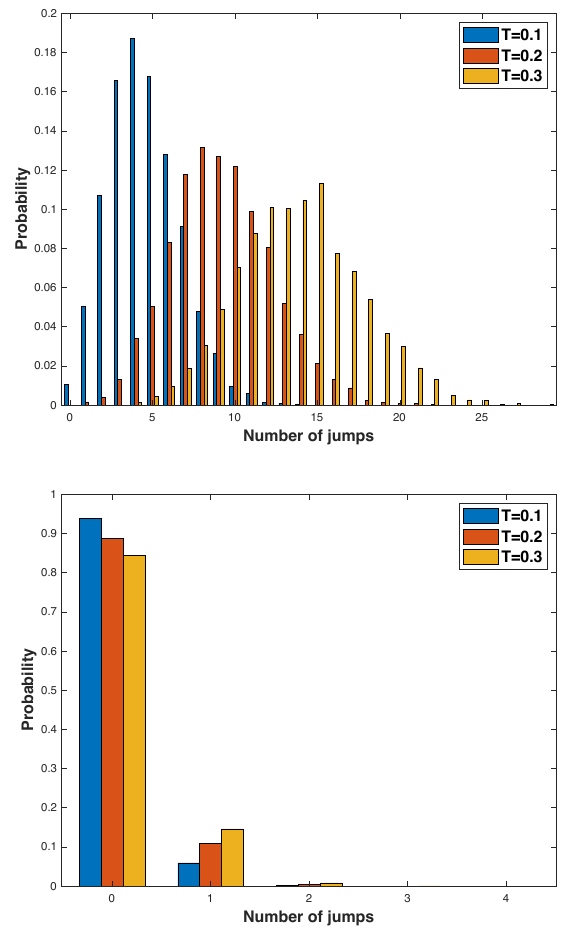}
			\caption{Type II: Probability of jumps for different values of $T$, when $\Delta=1$, $\gamma=1$, $\theta=\frac{\pi}{2}$ and $v=0.1$ (top) $v=10$ (bottom).}
		\label{fig:T2_PvsNDiffTThetaPi2}
	\end{figure}
	\begin{figure}
		\centering
		\includegraphics[width=\columnwidth]{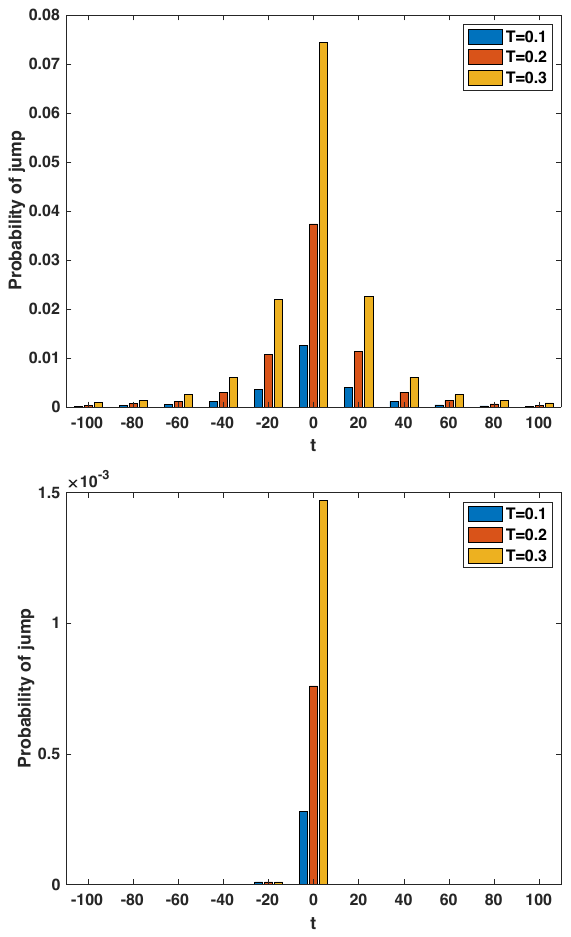}
			\caption{Type II: Probability of jumps in time intervals $\Delta t=20$ for different values of $T$, when $\Delta=1$, $\gamma=1$, $\theta=\frac{\pi}{2}$ and $v=0.1$ (top), $v=10$ (bottom).}
		\label{fig:LZ2IntervalsGamma1ThetaPi2}
	\end{figure}


	\section{Statistics of quantum jumps }
	\label{sec:statistics}
	In this section, we discuss the results for two types of dissipative Landau-Zener models. In all figures the evolution starts at $t=-100$ to $t=100$ and $\Delta=1$. 
	\subsection{Type I}
	Here we report our results of the statistical information obtained by analyzing quantum trajectories for dissipative Landau-Zener type I. We present the role of temperature and coupling strength to the environment on the statistics of the jumps. In all figures on this subsection, the initial state is $\ket{g}$, the eigenstate of $\sigma_z$ with the negative eigenvalue. 
	
	Figure~(\ref{fig:T1_PvsNDiffGamma}) shows the probability of having different numbers of jumps when $v=0.1$ (top) and $v=10$ (bottom). In this figure $\Delta=1$ and $\tau=0$. 
	As it is seen in Fig.~(\ref{fig:T1_PvsNDiffGamma}), for  $v=0.1$ by increasing $\gamma$, coupling interaction with the environment, the median of the probability distribution shifts toward larger values which indicates larger numbers of jumps, as expected. Furthermore, by increasing $\gamma$ the probability distribution becomes more symmetric. 
	For large enough $\gamma$ such as $\gamma=0.8$ the mean, mode and median of the statistical distribution are all equal and we have a normal statistical distribution. 
	When $v=10$ and the evolution window like the previous case is from $t=-100$ to $t=100$, there is a low chance of having quantum jumps. Even increasing $\gamma$ does not lead to considerable change in the statistical distribution of the number of jumps. 
	
	The role of temperature in the statistical distribution has been represented in Fig.~(\ref{fig:T1_PvsNDiffTau}). In Fig.~(\ref{fig:T1_PvsNDiffTau}), for $\gamma=0.1$ and $\Delta=1$, we have the statistics of quantum jumps for different values of $\tau$. For both $v=0.1$ (top) and $v=10$ (bottom), by increasing $\tau$, the median of the distributions increases and the distribution becomes more symmetric. For large enough $\tau$, quantum jumps have normal statistical distribution.

	Quantum trajectories provide statistics not only of the total number of jumps but also of jumps in different time intervals. Figure~(\ref{fig:LZ1IntervalsTau0}) shows the quantum jump statistics in the time intervals $\Delta t=20$. In both subplots $\tau=0$. For $v=0.1$ (top), the jumps are distributed symmetrically around $t=0$, and the interval with the maximum probability is at $t=0$ when the system has its minimum gap. For $v=10$ (bottom), there is a small chance of having quantum jumps, and if there are any, they occur after $t=0$. Both subplots confirm that by increasing $\gamma$ more jumps occur in the time intervals $\Delta t$. 
	
	For a fixed value of $\gamma=0.1$ and $\Delta=1$, Fig.~(\ref{fig:LZ1IntervalsGamma01}) shows the quantum jump statistics in the time intervals with width $\Delta t$ for different values of $\tau$ when $v=0.1$ (top) and $v=10$ (bottom). For $v=0.1$ (top), we see that by increasing temperature, the probability distribution is again symmetric around $t=0$ like the case of $\tau=0$ represented in the top subplot of Fig.~(\ref{fig:LZ1IntervalsTau0}). 
	The difference is that the normal distribution at $\tau=0$ tends to uniform distribution by increasing $\tau$. For $v=10$, by fixing $\gamma=0.1$ and increasing $\tau$, the quantum jumps occur from the beginning of the evolution and time intervals get more populated. This is expected because at $\tau=0$ the only possible jump is a projection to $\ket{g}$ described by $C_1^{(\rm I)}$ in Eq.~(\ref{eq:JumpTI}). By increasing $\tau$, not only does the probability of jumps to $\ket{g}$ increase but also there is a possibility of jump described by the jump operator $C^{(\rm I)}_2=\sqrt{\tau}\sigma_+$ in Eq.~(\ref{eq:JumpTI}) Therefore, there are chances that any state gets projected to $\ket{e}$. By increasing $\tau$, these probabilities increase, which explains the statistics in Fig.~(\ref{fig:LZ1IntervalsGamma01}).
	
	\subsection{Type II}
	In this subsection, we report the results for the type II dissipative Landau-Zener model. Our aim is to see the role of temperature and spin-coupling direction on the statistics of jumps.

In all proceeding analysis, the initial state is the instantaneous eigenstate of $H_{\rm LZ}$ in Eq.~(\ref{eq:LZ-Hamiltonian}), at $t=-\infty$. When $T=0$, the only Lindblad operator contributing to the evolution is $C_1^{(\rm II)}$. Because the initial state at $t=-\infty$ is $\ket{\epsilon_-(-\infty)}$, the probability of jump $\delta p$ is zero. Therefore, the evolution of the state is governed by the Hamiltonian. By repeating the same argument for proceeding steps, we expect the whole evolution to be governed by the Hamiltonian, and the system remains in its instantaneous ground state. The statistical investigations represented in Fig.~(\ref{fig:T2_PvsNDiffTGammaT0Theta0V01}) confirm this argument. Fig.~(\ref{fig:T2_PvsNDiffTGammaT0Theta0V01}) shows the statistical distribution for $v=0.1$ and $\theta=0$. The same result is seen for other values of $v$ and $\theta$.

	Figure~(\ref{fig:T2_PvsNDiffTTheta0}) shows the statistics of number of jumps when $\lambda=1$, $\theta=0$ for $T=0.1, 0.2$ and $0.3$. In both cases of $v=0.1$ (top) and $v=10$ (bottom), the statistical distributions are very close to the normal distributions. 
	By increasing temperature $T$, the median and the width of the distributions become larger. 
	Furthermore, we see that for each fixed value of bath's temperature $T$, by changing $v=0.1$ to $v=10$, the mean and the width of the statistical distribution increases. 
	The quantum jumps distribution in time-intervals $\Delta t=20$ is presented in Fig.~(\ref{fig:LZ2IntervalsGamma1Theta0}). We see that although the statistical distributions of number of jumps for both $v=0.1$ and $v=10$ are similar, the distribution of jumps in time is completely different for $v=0.1$ and $v=10$. While for $v=0.1$ the distribution is symmetric around $t=0$, for $v=10$ after system approaches its minimum gap at $t=0$, the probability that a system experiences a quantum jump increases. 
	
	For transversal spin-coupling ($\theta=\frac{\pi}{2}$) the probability distributions of the number of jumps are shown in Fig.(\ref{fig:T2_PvsNDiffTThetaPi2}). In the top subplot ($v=0.1$) it is seen that by increasing the bath's temperature, the median of the probability distribution of the number of jumps increases. But for $v=10$ (bottom), the bath's temperature is not as effective as is when $v=0.1$. 
	
	By comparing the top (bottom) subplots of Fig.~(\ref{fig:T2_PvsNDiffTTheta0}) and Fig.~(\ref{fig:T2_PvsNDiffTThetaPi2}) we see that for the same set of parameters, the median of the probability distributions for longitudinal spin-coupling is larger than the median for transversal spin-coupling.
	
	The distribution of quantum jumps for transversal spin-coupling ($\theta=\frac{\pi}{2}$) in time intervals $\Delta t=20$ is shown in Fig.~(\ref{fig:LZ2IntervalsGamma1ThetaPi2}). When $v=0$, getting closer to the minimum gap ($t=0$), the probability of jumps increases, which is exactly the reverse for evolution with longitudinal spin-coupling as seen in the top subplot of Fig.~(\ref{fig:LZ2IntervalsGamma1Theta0}). As is seen in the bottom subplot of Fig.~(\ref{fig:LZ2IntervalsGamma1ThetaPi2}), the rare quantum jumps occure around $t=0$. 
	\section{Conclusion}\label{sec:Conclusion}
	We have studied statistics of quantum jumps in two types of dissipative Landau-Zener models. The Dissipative Landau-Zener model is a successful model for examining adiabatic/non-adiabatic transitions with significant applications in adiabatic quantum computation and quantum annealing in the presence of noise.  
	
	We have considered two types of dissipative Landau-Zener model. In type I, the jump operators project any state to the eigenstates of the Landau-Zener Hamiltonian at $t\to -\infty$. 
	In the type II dissipative Landau-Zener model we consider a particular form of interaction between the system and the environment which depends on the spin-coupling direction. In this model, the jumps operators are time-dependent and project any state to the instantaneous eigenstates of the Landau-Zener Hamiltonian. Comparing the statistics of quantum jumps in these two types of dissipative dynamics, reflects the role of quantum jumps. Furthermore, 
	comparing 
	Fig.~(\ref{fig:T2_PvsNDiffTTheta0}) with Fig.~(\ref{fig:T2_PvsNDiffTThetaPi2}) and comparing Fig.~(\ref{fig:LZ2IntervalsGamma1Theta0}) with (\ref{fig:LZ2IntervalsGamma1ThetaPi2}) shows the evident role of spin-coupling direction in statistics of quantum jumps. 
	
	The statistics reported here are obtained using the quantum trajectory approach, which studies the trajectory of pure states in time and the abrupt changes in them due to the noise. The average over all possible trajectories in time, gives the density operator of the system. Indeed, by examining the average description of the system, namely the density operator, it is not possible to retrieve the information about abrupt transitions in the system and obtain the statistics of jumps. That is while these statistics have practical importance in running quantum tasks. Here our focus was on a model applicable in adiabatic quantum computations. But this approach can be extended to various quantum tasks including quantum communications such as entanglement distribution and quantum network development.

	\section*{Acknowledgments}
	This work was supported by PNRR MUR project PE0000023- NQSTI, and by the European Union (ERC, RAVE, 101053159). Views and opinions expressed are however those of the author(s) only and do not necessarily reflect those of the European Union or the European Research Council. Neither the European  Union nor the granting authority can be held responsible. L.M. acknowledge financial support from the Iran National Science Foundation (INSF) under Project No. 4022322 and support from the ICTP through the Associates Programme (2019-2024).

	\bibliography{QTraj2024}
	
\end{document}